\documentclass[aps,prl,preprint,superscriptaddress]{revtex4}

\usepackage{graphicx}
\usepackage{dcolumn}
\usepackage{bm}

\begin{document}


\title{Optical Phase-Space-Time-Frequency Tomography}


\author{Paul Rojas, Rachel Blaser, Yong Meng Sua, and Kim Fook Lee}
\email[]{kflee@mtu.edu}
\affiliation{%
Department of Physics,\\ Michigan Technological
University,\\ Houghton, Michigan 49931}%


\date{\today}

\begin{abstract}
We present a new approach for constructing optical
phase-space-time-frequency tomography (OPSTFT) of an optical wave
field. This tomography can be measured by using a novel four-window
optical imaging system based on two local oscillator fields balanced
heterodyne detection. The OPSTFT is a Wigner distribution function
of two independent Fourier Transform pairs, i.e., phase-space and
time-frequency. From its theoretical and experimental aspects, it
can provide information of position, momentum, time and frequency of
a spatial light field with precision beyond the uncertainty
principle. Besides the distributions of $x-p$ and $t-\omega$, the
OPSTFT can provide four other distributions such as $x-t$, $p-t$,
$x-\omega$ and $p-\omega$. We will simulate the OPSTFT for a light
field obscured by a wire and a single-line absorption filter. We
believe that the four-window system can provide spatial and temporal
properties of a wave field for quantum image processing and
biophotonics.
\end{abstract}

\pacs{03.67.Hk, 42.50.Dv, 42.65.Lm}

\maketitle


\section{Introduction}
Most of optical imaging methods are limited by the temporal and
spatial resolutions because of the unwanted scattering light coupled
into their detection systems. This limitation is due to the
fundamental concept in the process of measurement, that is, the
uncertainty principle.  Better resolution in position will reduce
the resolution in momentum (angle) of an optical imaging system.
Similarly, better resolution in time domain will reduce the
flexibilities of spectroscopic analysis on a physical object.
According to the uncertainty principle, the position $x$ and
momentum $p$, time $t$ and frequency $\omega$ of a spatial light
field cannot be measured simultaneously with high resolution.
However, the distribution of $x$ and $p$, $t$ and $\omega$ of the
spatial light field can be measured simultaneously with high
resolution by using two local oscillator fields. The use of two
local oscillator fields in a balanced heterodyne detection scheme is
also called two-window technique~\cite{kim99, kim10}. The optical
phase-space tomography (OPST)~\cite{kim99, Reil05} is Wigner
distribution~\cite{Wigner32, Hillery84} associated with a Fourier
Transform pair of position and momentum (angle) coordinates. Wigner
distribution associated with two independent Fourier Transform pairs
such as position-momentum and time-frequency is called optical
phase-space-time-frequency tomography (OPSTFT) or Wigner
phase-space-time-frequency distribution. In this paper, we develop a
four-window heterodyne detection scheme based on two local
oscillator (LO) fields for measuring the OPSTFT. The OPSTFT,
$\mathcal{W}(x, p, \omega, t)$, will offer the correlation
information of $x$, $p$, $t$, and $\omega$ of a wave field through
the distributions of $x-p$, $\omega-t$, $x-t$, $p-t$, $x-\omega$,
and $p-\omega$, where the two variables are plotted by fixing the
other two variables.

In quantum optics, Wigner function is usually used to represent the
quantum mechanical wave function or quantum state of a physical
system because there is no quantum device can directly measure the
wave function.
Raymer~\cite{smithey93,lvovsky09,raymer94,mcalister95, smith05} has
pioneered the measurement of Wigner distributions for
quadrature-field amplitude of non-classical state by using optical
homodyne detection. The method involves tomographic inversion (Radon
transform) of a set of measured probability distributions of
quadrature amplitudes. The measurement method developed by Raymer
has also been used to measure the Wigner distribution for transverse
spatial state (spatial mode) in a single photon level~\cite{smith05}
and time-frequency domain~\cite{beck93} of an optical
electromagnetic field.

Recently, residual spatial fluctuation in terms of small
displacements and tilts (momentum) of a whole optical beam has been
observed and used to exhibit EPR entanglement~\cite{Lam08}. Quantum
imaging~\cite{Mikhail06} has played a central role in understanding
spatial fluctuations in quantum regions. Controlling these quantum
fluctuations can improve image resolution and beam positioning for
targeting technology. Entanglement with a large number of modes such
as images has been accomplished by using four-wave mixing in an
atomic vapor~\cite{Boyer08}. The property of multimode squeezed
light ~\cite{Mikhail99} which allows us to increase the sensitivity
beyond the shot-noise limit, could create many interesting new
applications in optical imaging, high-precision optical
measurements, optical communications and optical information
processing. The Wigner function contains sub-Planck phase-space
structures~\cite{Zurek01, Zurek06}, which can be used to detect
small movement of a large object.

The main advantage of applying Wigner distribution in optical
imaging and sensing is certain wave-particle features of Wigner
distributions, related to optical coherence, can survive over
distances that are large compared to the transport mean free
path~\cite{John96}. Another advantage is that Wigner phase-space
distributions could bridge the gap between phenomenological
transport equations and rigorous wave equation treatments. Since
rigorous transport equations can be derived for Wigner
distributions, they are essential for obtaining fundamental new
insights into the nature of light propagation in multiple scattering
media. Evolution equations for Wigner distributions, which include
optical coherence scatterings, are generally non-local and are
relatively unexplored. With suitable approximations, these non-local
equations reduce to the usual radiative transport
equations~\cite{Ishimaru78}. Establishing the physical relationship
between Wigner distributions and the phenomenological specific
intensity will impact most existing methods of imaging in multiple
scattering media.

For a wave field varying in one spatial dimension and one spectra
domain, $\mathcal{E}(x,\omega)$, the Wigner
phase-space-time-frequency distribution is given by,
\begin{equation}
\mathcal{W}(x,p, \omega, t)=\int \frac{d\epsilon}{2\pi}e^{i\epsilon
p} \int \frac{d\Omega}{2\pi}e^{i\Omega t}
\langle\mathcal{E}^{*}(x+\frac{\epsilon}{2},
\omega+\frac{\Omega}{2})\mathcal{E}(x-\frac{\epsilon}{2},
\omega-\frac{\Omega}{2})\rangle \label{eq:01}
\end{equation}
where $\langle...\rangle$ denotes a statistical average. It is easy
to show that $\int dp\int dt\mathcal{W}(x, p, \omega,
t)=|\mathcal{E}(x,\omega)|^{2}$, $\int dx \int d\omega
\mathcal{W}(x,p, \omega, t)=|E(p,t)|^{2}$, and etc. Most important,
Eq.~\ref{eq:01} shows that the Wigner distribution is Fourier
Transform related to the two-point mutual coherence function in
position and spectra. Therefore, it is sensitive to the spatially
and spectrally varying phase and amplitude of the field.

\section{Theoretical Approach: Measurement of OPSTFT}

In this work, we develop a four-window heterodyne detection scheme
based on two local oscillator (LO) fields for measuring the OPSTFT.
\begin{figure}
\includegraphics[scale=1.0]{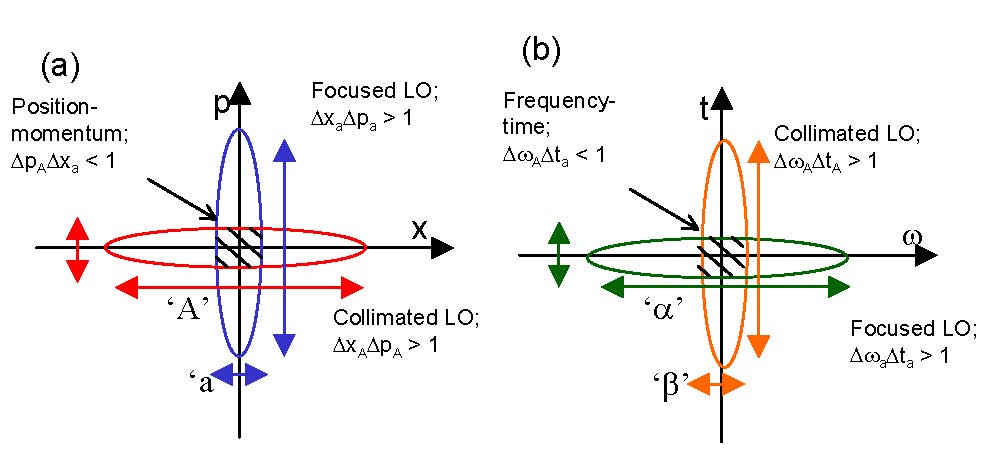}
\caption{(a) the x-p and (b)$\omega$-t representations of the LO
fields. The shaped area is position-momentum and time-frequency
resolutions showing beyond the uncertainty limits.}
\end{figure}
As shown in Fig.1(a) and (b), a local oscillator (LO) field is a
phase coherent superposition of two fields i.e a focused and a
collimated Gaussian fields. The focused LO Gaussian beam has spatial
width of $\Delta x_{a} = a$ and broad optical spectrum with
bandwidth of $\Delta \omega_{a} = \alpha$. The collimated (large) LO
Gaussian beam has spatial width of $\Delta x_{A} = A$ and narrow
optical spectrum with bandwidth of $\Delta \omega_{A} = \beta$. The
position resolution $a$ is provided by the focused LO Gaussian beam.
The momentum resolution $1/A$ is provided by the collimated LO
Gaussian beam. The purpose of this arrangement is to obtain
independent control of position and momentum (angle) resolution such
that the product of $(\Delta x_{a}\cdot \Delta p_{A}) = a \cdot 1/A
\leq 1$ clearly surpasses the uncertainty principle limit as shown
in the shaped area in Fig.1(a). Simultaneously, the method can
provide independent control of frequency (spectra) and time
(path-length) resolution such that $(\Delta \omega_{A} \cdot\Delta
t_{a}) = \beta \cdot 1/\alpha \leq 1$ to surpass the uncertainty
principle limit as shown in the shaped area in Fig.1(b). The
spectra-resolved resolution $\beta$ is provided by the narrowband
collimated LO Gaussian beam. The path-resolved resolution $1/\alpha$
is provided by the broadband focused LO Gaussian beam. In other
words, we use position (time) window to obtain position (temporal)
information of an optical field and simultaneously use momentum
(frequency) window to reject noises due to other propagating angles
(frequencies) from coupling into the detection system.
\begin{figure}
\includegraphics[scale=1.0]{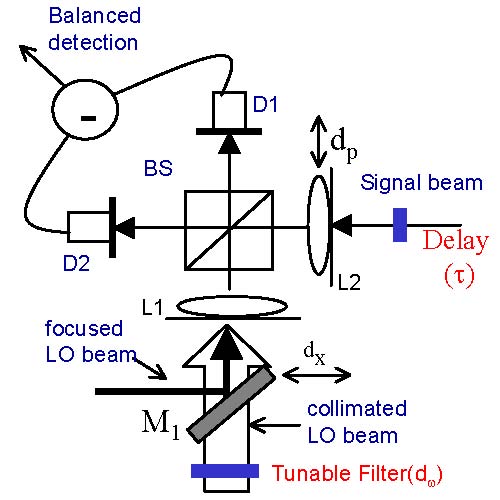}
\caption{Four-window balanced heterodyne detection for measuring
OPSTFD.}
\end{figure}

A schematic setup of the four-window technique is shown in Fig.2. In
this scheme, optical phase-space-time-frequency tomography (OPSTFT)
is measured by scanning transverse position $(d_{x})$ and transverse
momentum $(d_{p})$, and by tuning wavelength $(d_{\omega})$ and path
delay $(\tau)$. Our system employs balanced heterodyne detection of
the probe signal field, which is overlapped with two strong local
oscillator fields (LO). The beat amplitude $V_{B}$ is determined by
the spatial and spectra overlapping of two local oscillator (LO) and
signal (S) fields at the plane of the detector $(Z=Z_{D})$ as,
\begin{equation}
V_{B}=\int dx'\int d\omega
E^{\ast}_{LO}(x',\omega,z_{D})E_{S}(x',\omega,z_{D}) \label{eq:02}
\end{equation}
where $E_{LO(S)}(x',\omega,z_{D})$ is used to represent the two
local oscillator fields (signal field), respectively.  The $x'$
denotes the transverse position at the detector plane. As shown in
Fig.2, when the LO fields are translated off-axis for a distance of
$d_{x}$ and the tunable filter is tuned $d_{\omega}$ away from
optical center frequency, the LO field will have its spatial and
spectra arguments shifted in Eq.~\ref{eq:02} as given by,
\begin{equation}
V_{B}(d_{x}, d_{\omega})=\int dx'\int d\omega
E^{\ast}_{LO}(x'-d_{x},\omega -d_{\omega}, z_{D})E_{S}(x', \omega,
z_{D}). \label{eq:03}
\end{equation}
Note that the center frequency of the collimated LO beam is needed
to be tuned because we assume the focussed LO beam has a broad
optical spectrum. Using the Fresnel approximation and standard
Fourier optics technique, we can relate the fields at the detector
plane, $E(x', \omega, z_{D})$, to the fields at the source planes,
$E(x, \omega, z=0)$, of lenses L1 and L2 as follow; (i) The LO and
signal fields each will experience a spatially varying phase of
$e^{-\frac{ik x^2}{2f}}$ after passing through the lens. (ii) Since
the lens L2 in the signal beam is translated off-axis by $d_{p}$ and
the path-length of the signal beam is delayed by $\tau$, the signal
field will experience additional phase-shift of
$e^{-\frac{ik(x-d_{p})^{2}}{2f}}$ and delay of $e^{-i\omega\tau}$,
respectively. (iii) The LO and signal fields propagate to the
detector plane through a distance of $d=f$, which is the
focus-length of lenses L1 and L2, and will experience the
phase-shift of $e^{-\frac{ik(x-x')^{2}}{2f}}$. From (i), (ii) and
(iii), the LO and signal fields in Eq.~\ref{eq:03} can be rewritten
in terms of the input field $z=0$ as described in~\cite{kim10},
\begin{eqnarray}
E_{LO}(x'-d_{x}, \omega-d_{\omega}, z_{D})=\sqrt{\frac{k}{i 2\pi
f}}\int dx e^{i\frac{k (x-x')^2}{2f}} e^{-i\frac{k x^2}{2 f}}
E_{LO}(x-d_{x}, \omega-d_{\omega}, z=0),\label{eq:04}
\end{eqnarray}
\begin{eqnarray}
E_{S}(x', \omega, z_{D}) =\sqrt{\frac{k}{i 2\pi f}}\int dx
e^{i\frac{k (x-x')^2}{2f}} e^{-i\frac{k (x-d_{p})^2}{2 f}}
e^{-i\omega \tau} E_{S}(x, \omega, z=0). \label{eq:05}
\end{eqnarray}
By using simple algebra, one obtains the mean square beat amplitude
as given by,
\begin{eqnarray}
|V_{B}(d_{x}, d_{\omega})|^{2}&=&\int d\omega \int dx
E^{*}_{LO}(x-d_{x}, \omega-d_{\omega}) E_{S}(x, \omega) e^{-i\frac{k
d_{p} x}{f}} e^{-i \omega \tau}\nonumber\\
&\times& \int d\omega' \int dx' E_{LO}(x'-d_{x}, \omega'-d_{\omega})
E^{*}_{S}(x', \omega') e^{i\frac{k d_{p} x'}{f}} e^{i \omega'
\tau}\label{eq:06}
\end{eqnarray}
Then, using the following variable transformations,
$x=x_{\circ}+\frac{\eta}{2}$,
$\omega=\omega_{\circ}+\frac{\eta_{\omega}}{2}$,
$x'=x_{\circ}-\frac{\eta}{2}$, and
$\omega'=\omega_{\circ}-\frac{\eta_{\omega}}{2}$, where the Jacobian
of this transformation is 1. The mean square heterodyne beat signal
of Eq.(6) can be rewritten in terms of these variables as,
\begin{eqnarray}
|V_{B}|^{2}&=&\int d\omega_{\circ} d\eta_{\omega} dx_{\circ} d\eta
E^{*}_{LO}(x_{\circ}-d_{x}+\frac{\eta}{2},
\omega_{\circ}-d_{\circ}+\frac{\eta_{\omega}}{2})
E_{LO}(x_{\circ}-d_{x}-\frac{\eta}{2},\omega_{\circ}-d_{\omega}-\frac{\eta_{\omega}}{2})\nonumber\\
&\times&
E^{*}_{S}(x_{\circ}-\frac{\eta}{2},\omega_{\circ}-\frac{\eta_{\omega}}{2})
E_{S}(x_{\circ}+\frac{\eta}{2},
\omega_{\circ}+\frac{\eta_{\omega}}{2}) e^{-i\frac{k d_{p} \eta}{f}}
e^{-i \eta_{\omega} \tau}.\label{eq:07}
\end{eqnarray}
Recall that the Wigner function is the Fourier Transform of the
two-point coherence function. Thus, the inverse transform for the
signal field is given by,
\begin{equation}
E^{*}_{S}(x_{\circ}+\frac{\eta}{2},
\omega_{\circ}+\frac{\eta_{\omega}}{2})E_{S}(x_{\circ}-\frac{\eta}{2},
\omega_{\circ}-\frac{\eta_{\omega}}{2})=\int dp dt e^{-i \eta p}
e^{-i \eta_{\omega} t} \mathcal{W}_{S}(x_{\circ}, p, \omega_{\circ},
t).\label{eq:08}
\end{equation}
By substituting Eq.(8) into Eq.(7), the mean square heterodyne beat
signal is then rewritten by,
\begin{eqnarray}
|V_{B}|^{2}&=&\int d\omega_{\circ} dx_{\circ} dp dt d\eta_{\omega}
d\eta E^{*}_{LO}(x_{\circ}-d_{x}+\frac{\eta}{2},
\omega_{\circ}-d_{\omega}+\frac{\eta_{\omega}}{2})E_{LO}(x_{\circ}-d_{x}-\frac{\eta}{2},
\omega_{\circ}-d_{\omega}-\frac{\eta_{\omega}}{2})\nonumber\\
&\times& \mathcal{W}(x_{\circ}, p, \omega_{\circ}, t)
e^{-i\eta(\frac{kd_{p}}{f}+p)}e^{-i\eta_{\omega}(\tau+t)}
.\label{eq:09}
\end{eqnarray}
Using a similar procedure for the Wigner function of the LO field as
given by,
\begin{eqnarray}
\mathcal{W}_{LO}(x_{\circ}-d_{x}, p+k\frac{d_{p}}{f},
\omega_{\circ}-d_{\omega}, t-\tau) &=&\int d\eta
d\eta_{\omega}e^{i\eta (p+\frac{k
d_{p}}{f})}e^{i\eta_{\omega}}(t-\tau)\nonumber\\
&\times& E^{*}_{LO}(x_{\circ}-d_{x}, p+k\frac{d_{p}}{f},
\omega_{\circ}-d_{\omega}+\frac{\eta}{2})\nonumber\\
&\times&
E_{LO}(x_{\circ}-d_{x}-\frac{\eta}{2}-\omega_{\circ}-d_{\omega}-\frac{\eta_{\omega}}{2}).\label{eq:10}
\end{eqnarray}

Then, the mean square heterodyne beat signal is finally obtained as
given by,
\begin{equation}
|V_{B}|^{2}=\int dx \int d\omega \int dp \int dt
\mathcal{W}_{LO}(x-d_{x}, p+k\frac{d_{p}}{f}, \omega-d_{\omega},
t-\tau) \mathcal{W}_{S}(x, p, \omega, t)) \label{eq:11}
\end{equation}
where we have changed the notations of $x_{\circ}\rightarrow x$ and
$\omega_{\circ}\rightarrow \omega$. The $|V_{B}|^{2}$ is
proportional to the phase-space-frequency-time convolution integral
of the Wigner distributions for the two local oscillator and the
signal fields in the planes of the input lenses L1 and L2,
respectively.

We will use a broadband light source for this experiment. The LO
fields are engineered in the form as given by,
\begin{eqnarray}
E_{LO}(x, \omega)=
E_{o}[\textrm{exp}(-\frac{x^{2}}{2a^{2}})\textrm{exp}(-\frac{\omega^{2}}{2\alpha^{2}}))+\gamma
\textrm{exp}(-\frac{x^{2}}{2A^{2}})\textrm{exp}(-\frac{\omega^{2}}{2\beta^{2}})e^{i\phi}]
\label{eq:12}
\end{eqnarray}
where the spectrum of LO fields are assumed to be Gaussian function.
This can be accomplished by using a single mode fiber with a tunable
bandpass Gaussian filter from Newport. The phase- dependent part of
Wigner function for the LO takes the form
\begin{eqnarray}
W_{LO}(x, p, \omega, t)&\propto&
\textrm{exp}[\frac{2x^{2}}{A^{2}}-2a^{2}p^{2}+\frac{2\omega^{2}}{\alpha^{2}}-2\beta^{2}t^{2}]\cos(2xp+2\omega
t+\phi)\nonumber\\
&\cong&\cos(2xp+2\omega t+\phi)\label{eq:13}
\end{eqnarray}
where we take the approximation of $A\gg a$ and $\alpha \gg \beta$.
Then, the range of integration for the momentum, position, frequency
and time coordinates in Eq.~\ref{eq:11} is limited by the signal
field. In this scheme (similar electronic components as in
ref[~\cite{kim99,kim10}]), the two LO fields which differ in
frequency by $\phi$ kHz are phase-locked. The signal field is
modulated such that the heterodyne beat signals with the focused LO
field and the collimated LO field are about $\Omega_{\omega}$ MHz
and $\Omega_{\omega}$ MHz + $\phi$ kHz, respectively. The root mean
square beat amplitude is measured with an analog spectrum analyzer
with a bandwidth of 100 kHz ($>\phi$ kHz) centered at
$\Omega_{\omega}$ MHz. The output of the spectrum analyzer is
squared in real time with a low noise amplifier and an analog
multiplier, then the amplified signal is sent to the
lock-in-amplifier. Substituting Eq.~\ref{eq:13} into
Eq.~\ref{eq:11}, we find that the in- and out-of phase quadrature
amplitudes in the lock-in-amplifier are directly corresponding to
the real and imaginary parts of the quantity,
\begin{eqnarray}
|V_{B}|^{2}&\propto& \mathcal{K}(x_{o}, p_{o}, \omega_{o}, t_{o})\nonumber\\
&\propto& \int
dx'dp'd\omega'dt'e^{[2i(x'-x_{o})(p'-p_{o})+2i(\omega'-\omega_{o})(t'-t_{o})]}\mathcal{W}_{S}(x',p',
\omega', t')\nonumber\\
&\propto& \langle E^{*}_{S}(x_{o},
\omega_{o})E_{S}(p_{o},t_{o})\rangle exp(i x_{o} p_{o}+ i \omega_{o}
t_{o})\nonumber\\
&\propto& S_{R}+i S_{I} \label{eq:14}
\end{eqnarray}
The $\mathcal{K}$(x, p, $\omega$, t) is the Kirkwood-Rihaczek
phase-space-time-frequency distribution. Eq.~\ref{eq:14} is readily
inverted to yield the Wigner phase-space and time-frequency function
or the OPSTFT of the signal field by a linear transformation. We
obtain,
\begin{eqnarray}
W_{S}(x, p, \omega, t) &\propto& \int
dx_{\circ}dp_{\circ}d\omega_{\circ} dt_{\circ}
\cos[2(x-x_{o})(p-p_{o})+2(\omega-\omega_{o})(t-t_{o})]S_{R}(x_{\circ},p_{\circ},
\omega_{\circ}, t_{\circ})\nonumber\\
&+& \int dx_{\circ}dp_{\circ}d\omega_{\circ}
dt_{\circ}\sin[2(x-x_{o})(p-p_{o})+2(\omega-\omega_{o})(t-t_{o})]S_{I}(x_{\circ},p_{\circ},
\omega_{\circ}, t_{\circ})\nonumber\\ \label{eq:15}
\end{eqnarray}
which is the OPSTFT of the signal field. $S_{R}$ and $S_{I}$ are the
real and imaginary parts of Eq.~\ref{eq:14}, i.e. the in- and
out-of-phase quadrature amplitudes, which are simultaneously
measured.

\section{Simulation of OPSTFT}
\begin{figure}
\includegraphics[scale=0.5]{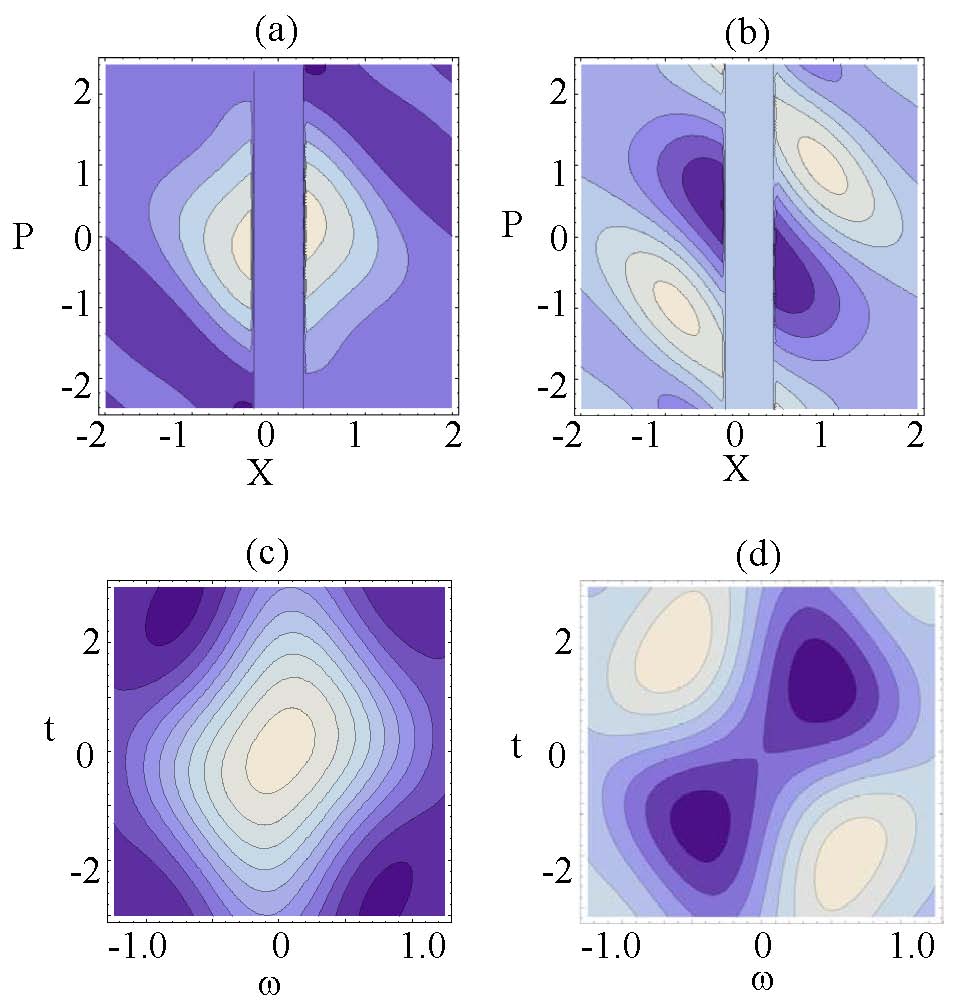}
\caption{The Kirkwood-Rihaczek ($\mathcal{K}$) distribution (a) real
and (b) imaginary parts of $\mathcal{K}(x, p, 0, 0)$. (c) real and
(d)imaginary parts of $\mathcal{K}(0.4, 0, \omega, t)$ }
\end{figure}

\subsection{A Thin Wire}

 The measurement of OPSTFT based on two local oscillator
fields is called four-window technique. In practice, the four
independent variables $a$, $1/A$, $1/\alpha$ and $\beta$ are chosen
to be small to resolve scales of interest on the spatial and
temporal properties of a wave field. Physical properties of an
object can be extracted through measuring the OPSTFT of the
scattered light field through the object. The OPSTFT can provide
$x-p$ and $\omega-t$ distributions, and other four distributions
such as $\omega-x$, $t-x$, $\omega-p$ and $t-p$ distributions. These
six distributions can provide new types of information for the light
field under study. We numerically simulate the measurement of OPSTFT
for the light field scattered through a thin wire with the diameter
of 0.6 mm. We use a Gaussian beam with a wave field as given by,
\begin{equation}
\mathcal{E}(x,t)\propto
\textrm{exp}[-\frac{x^2}{2\sigma_{x}^{2}}]\textrm{exp}[i\frac{k
x^2}{2R}]\textrm{exp}[-\frac{t^2}{2\sigma_{t}^{2}}]\textrm{exp}[i\omega_{\circ}t_{\circ}],\label{eq:16}
\end{equation}
where $\sigma_{x}$ and $\sigma_{t}$ are the spatial and temporal
bandwidths. R and $\omega_{\circ}$ are the radius of curvature and
center/carrier frequency of the light field. We are interested in
looking at the light scattered right after the wire, where the
scattered light field, $\mathcal{E}_{wire}(x, t)$, can be written as
the product of Eq.~\ref{eq:16} and a wire function
(slitfun[x]=If[$-0.3\textrm{mm}\leq x \leq 0.3\textrm{mm}$, 0.0,
1.0] in Mathematica program). We have used this field for exploring
phase-space interference analog to superposition of two spatially
separated coherent states~\cite{kim10}. However, the phase-space
distribution associated with time-frequency distribution
incorporated into Wigner distributions haven't been explored. We use
$\sigma_{x}$ = 0.85 mm, $\sigma_{t}$=200 fs, and $R$=-10000 mm and
$\omega_{\circ}=0$ for the simulation. We first obtain
$\mathcal{E}_{wire}(p,\omega)$ by numerically Fourier transforming
the $\mathcal{E}_{wire}(x, t)$. Then, we generate the
Kirkwood-Richaczek phase-space-time-frequency distribution,
$\mathcal{K}(x, p, \omega,
t)=\mathcal{E}^{*}_{wire}(x,t)\mathcal{E}_{wire}(p,
\omega)\textrm{exp}(i x p - i \omega t)$, as in Eq.~\ref{eq:14} for
the scattered field $\mathcal{E}_{wire}(x, t)$.
\begin{figure}
\includegraphics[scale=0.5]{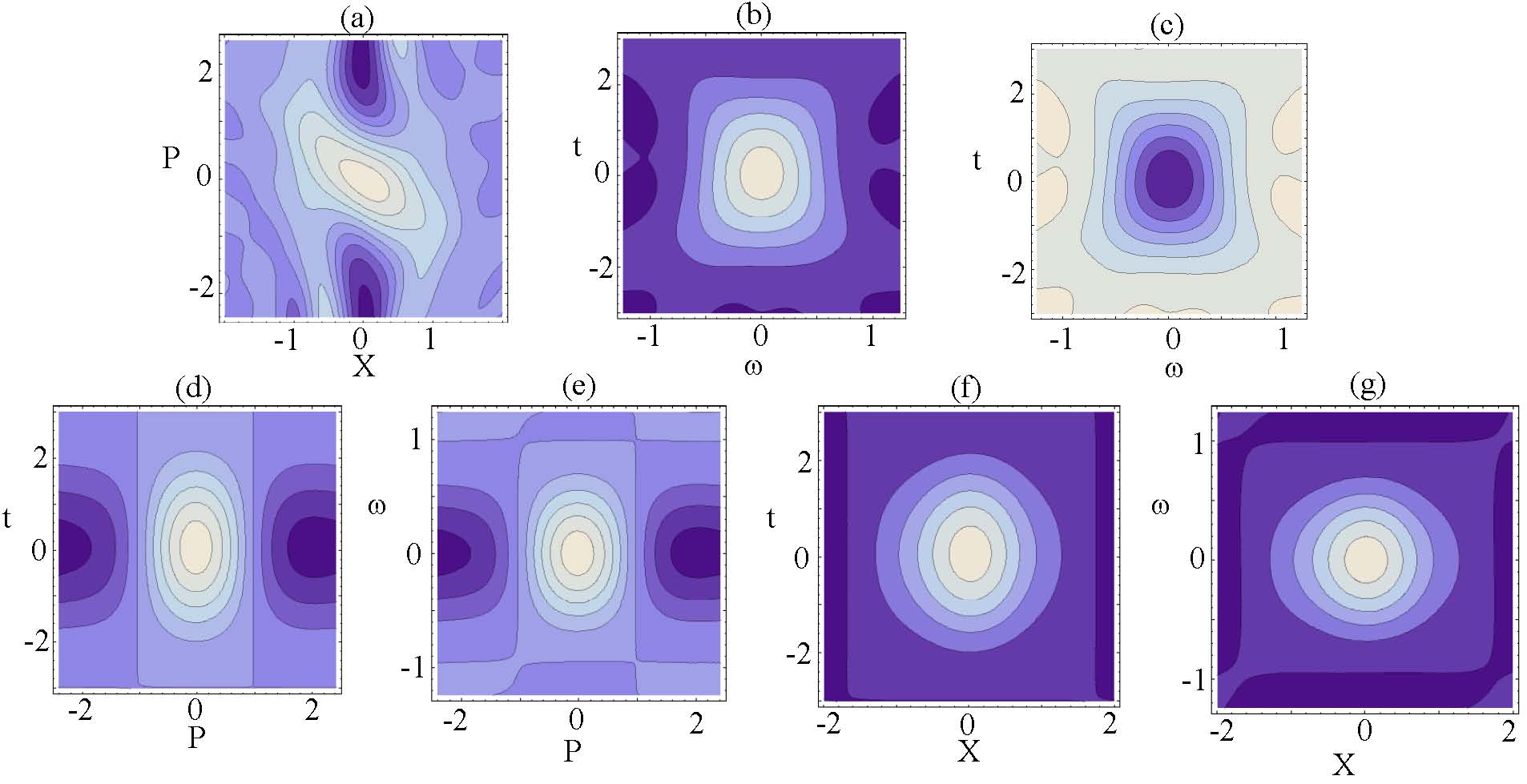}
\caption{The optical phase-space-time-frequency tomography (OPSTFT)
(a)$\mathcal{W}(x, p, 0, 0)$, (b)$\mathcal{W}(0, 0, \omega, t)$,
(c)$\mathcal{W}(0, 2, \omega, t)$, (d)$\mathcal{W}(0, p, 0, t)$,
(e)$\mathcal{W}(0, p, \omega, 0)$, (f) $\mathcal{W}(x, 0, 0, t)$,
and $\mathcal{W}(x, 0, \omega, 0)$ .}
\end{figure}
We first plot the real and imaginary parts of $\mathcal{K}(x, p, 0,
0)$ as shown in Fig.3(a) and (b). We adopt the units for position
(x) in mm, momentum (p) in rad/mm, frequency ($\omega$) in $10^{13}$
Hz, and time in $10^{-13}s$. The $\mathcal{K}(0, 0, \omega, t)$ is
zero because there is zero beat signal ($|V_{B}|^2$) at the position
x=0 and p =0. To explore the time-frequency distribution associated
with nonzero beat signal ($|V_{B}|^2$) at the phase-space
point(x=0.4, p=0), we plot the real and imaginary parts of
$\mathcal{K}(0.4, 0, \omega, t)$ as shown in Fig.3(c) and (d). From
the $\mathcal{K}(x, p, \omega, t)$ distribution, we can obtain
Wigner phase-space-time-frequency distribution by using linear
transformation as in Eq.~\ref{eq:15}. We use Mathematica program for
performing the numerical transformation. Fig.4(a) is the plot of
$\mathcal{W}(x, p, 0, 0)$ where the tunable filter in the collimated
LO beam is set at center frequency and the delay $(\tau)$ in the
signal beam is set to zero. The phase-space oscillation along the
x=0 is due to the coherent phase-space interference of two spatially
separated wave packets after the wire. Now, we plot the
$\mathcal{W}(0, 0, \omega, t)$ as shown in Fig.4(b), where the
phase-space point is at (0, 0). Note that the $\mathcal{K}(0, 0,
\omega, t)$ is zero everywhere but not the Wigner function of
$\mathcal{W}(0, 0, \omega, t)$. The reason is $\mathcal{K}$
distribution is suitable for describing local properties of a wave
function, while the Wigner function is suitable for describing wave
properties of a particle wave function~\cite{kim10}. Fig.4(c) shows
an interesting result of $\mathcal{W}(0, 2, \omega, t)$, which is
negative, i.e, the inverse of Fig.4(b). This is because the Wigner
distribution of $\mathcal{W}(0, 2, 0, 0)$ has negative value. These
properties are important to explore hyper-entanglement of intrinsic
properties of single-photon spatial qubit states. Other four
distributions such as $\mathcal{W}(0, p, 0, t)$, $\mathcal{W}(0, p,
\omega, 0)$, $\mathcal{W}(x, 0, 0, t)$, and $\mathcal{W}(x, 0,
\omega, 0)$ are plotted as shown in Fig.4(d), (e), (f) and (g),
respectively. The $\mathcal{W}(0, p, 0, t))$ and $\mathcal{W}(0, p,
\omega, 0)$ exhibit oscillation/interference behavior along momentum
coordinate. The phase-space interference due to the two spatially
separated wave packets influenced the distribution of momentum/angle
in the time and spectra domains.

\subsection{An Absorption Filter}
\begin{figure}
\includegraphics[scale=0.5]{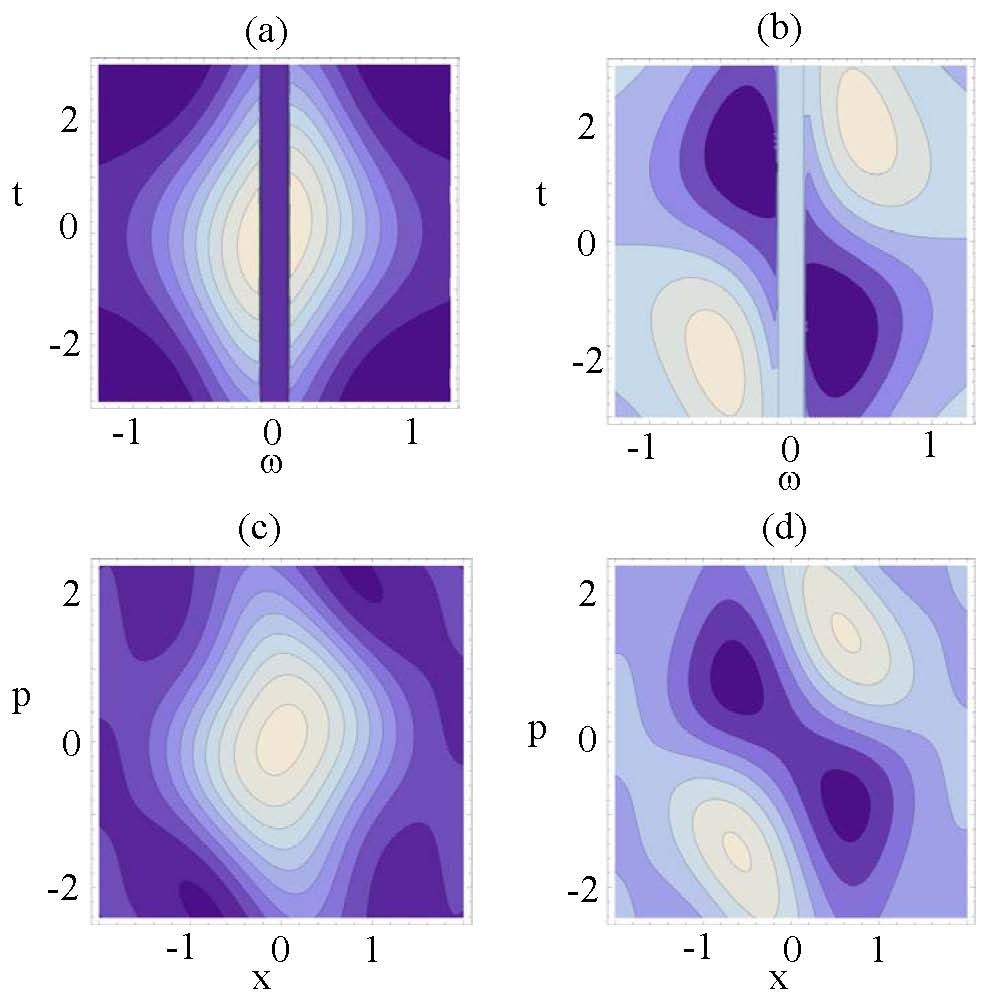}
\caption{The Kirkwood-Rihaczek ($\mathcal{K}$) distribution (a) real
and (b) imaginary parts of $\mathcal{K}(0, 0, \omega, t)$. (c) real
and (d)imaginary parts of $\mathcal{K}(x, p, 0.2, 0)$ }
\end{figure}
We numerically simulate the measurement of OPSTFT for a broadband
light field passing through an absorption filter with bandwidth of 2
THz. We use a Gaussian beam with a Gaussian linewidth as given by,
\begin{equation}
\mathcal{E}(x,\omega)\propto
\textrm{exp}[-\frac{x^2}{2\sigma_{x}^{2}}]\textrm{exp}[i\frac{k
x^2}{2R}]\textrm{exp}[-\frac{(\omega-\omega_{\circ})^2}{2
\sigma^{2}_{\omega}}],\label{eq:17}
\end{equation}
where $\sigma_{\omega}$ is the spectra bandwidth. It is much easier
to work on the spectra domain of the field so that the light field
passing through the filter can be written as the product of
Eq.~\ref{eq:17} and an absorption filter function
(slitfun[x]=If[$-0.1\leq x \leq 0.1 $, 0.0, 1.0] in Mathematica
program). We use $\sigma_{x}$ = 0.85 mm, $\sigma_{t}$=5.0 THz,
$R$=-10000 mm and $\omega_{\circ}=0$ for the simulation. First, we
generate the Kirkwood-Richaczek phase-space-time-frequency
distribution, $\mathcal{K}(x, p, \omega,
t)=\mathcal{E}^{*}_{filter}(x,\omega)\mathcal{E}_{filter}(p,t)\textrm{exp}(i
x p + i \omega t)$, where the $\mathcal{E}_{filter}(p, t)$ is
obtained by numerically Fourier transformed the
$\mathcal{E}_{filter}(x, \omega)$.
\begin{figure}
\includegraphics[scale=0.5]{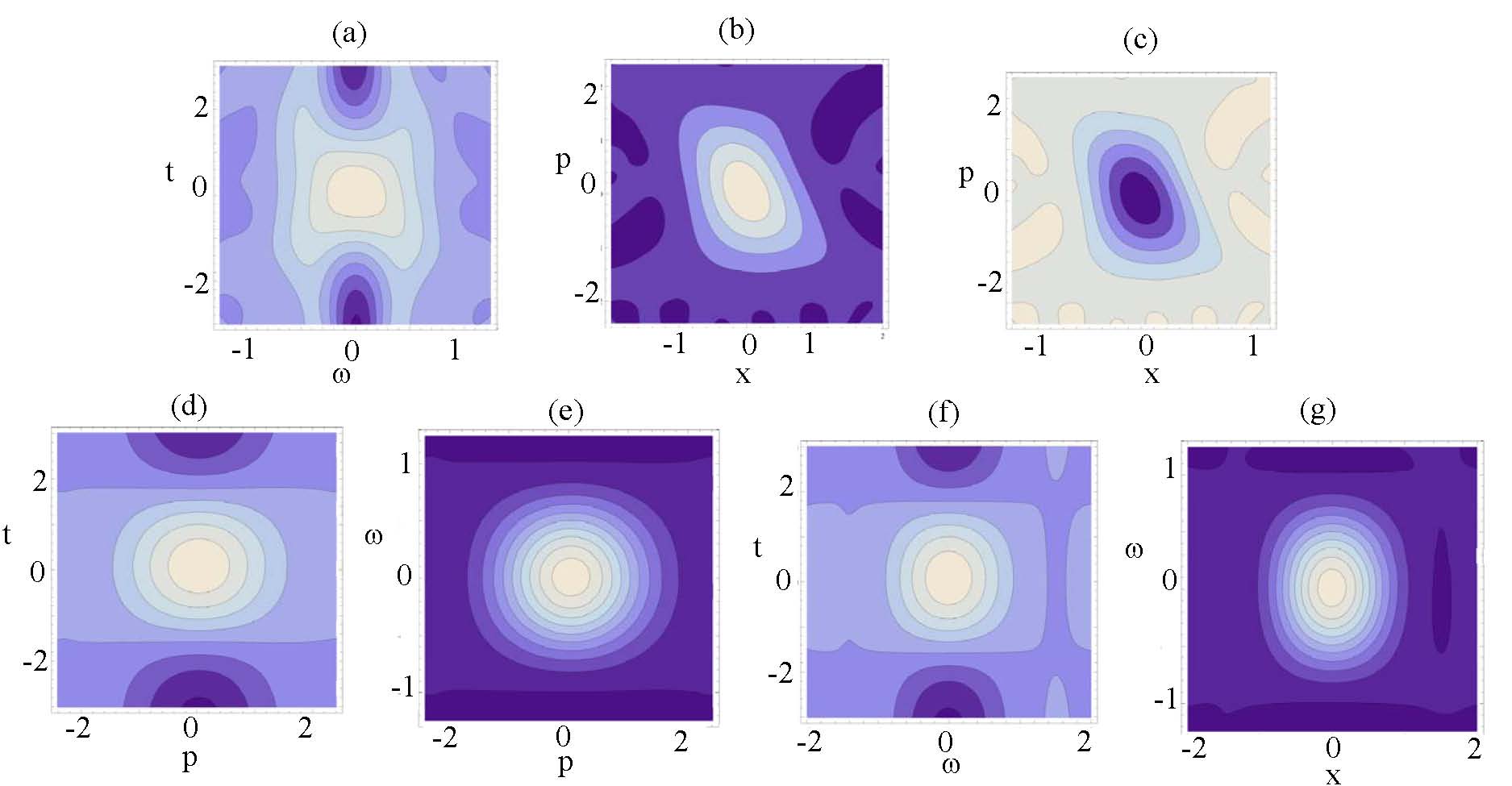}
\caption{The optical phase-space-time-frequency tomography (OPSTFT)
(a)$\mathcal{W}(0, 0, \omega, t)$, (b)$\mathcal{W}(x, p, 0, 0)$,
(c)$\mathcal{W}(x, p, 0, 3)$, (d)$\mathcal{W}(0, p, 0, t)$,
(e)$\mathcal{W}(0, p, \omega, 0)$, (f) $\mathcal{W}(x, 0, 0, t)$,
and $\mathcal{W}(x, 0, \omega, 0)$ .}
\end{figure}
The $\mathcal{K}(x, p, 0, 0)$ is zero because there is zero beat
signal ($|V_{B}|^2$) at the position $\omega=0$ and $t =0$. Fig.5(a)
and (b)show the real and imaginary parts of $\mathcal{K}(0, 0,
\omega, t)$. To explore the position-momentum distribution
associated with nonzero beat signal ($|V_{B}|^2$) at the
time-frequency point($\omega$=0.2, t=0), we plot the real and
imaginary parts of $\mathcal{K}(0.2, 0, \omega, t)$ as shown in
Fig.5(c) and (d). Since we have the $\mathcal{K}(x, p, \omega, t)$
distribution, then we can obtain Wigner phase-space-time-frequency
distribution by using linear transformation as in Eq.~\ref{eq:15}.

Fig.6(a) is the plot of $\mathcal{W}(0, 0, \omega, t)$ where the
position $d_{x}$ of the mirror for the LO beam is set to zero and
the momentum $d_{p}$ of the lens L1 in the signal beam is set to
zero. The time-frequency oscillation along the $\omega=0$ is due to
the coherent time-frequency interference of two spectrally separated
wave packets after the filter. This observation has been observed in
Wigner time-frequency distribution~\cite{beck93}. We plot the
$\mathcal{W}(x, p, 0, 0)$ as shown in Fig.6(b), where the
time-frequency point is at (0, 0). Fig.6(c) shows an interesting
result of $\mathcal{W}(x, p, 0, 3)$, which is negative, i.e, the
inverse of Fig.6(b). This is because the Wigner distribution of
$\mathcal{W}(0, 0, 0, 3)$ has negative value. Other four
distributions such as $\mathcal{W}(0, p, 0, t)$, $\mathcal{W}(0, p,
\omega, 0)$, $\mathcal{W}(x, 0, 0, t)$, and $\mathcal{W}(x, 0,
\omega, 0)$ are plotted as shown in Fig.6(d), (e), (f) and (g),
respectively. The $\mathcal{W}(0, p, 0, t))$ and $\mathcal{W}(x, 0,
0, t)$ exhibit oscillation/interference behavior along p =0 and x=0,
respectively.

\section{Conclusion}

In conclusion, we have developed four-window optical heterodyne
imaging technique based on two local-oscillator (LO) beams for
measuring OPSTFT of a wave field. The four-window technique can
simultaneously provide high resolution in position, momentum
(angle), time, and spectra for characterizing spatial properties of
a wave function. This newly developed optical
phase-space-time-frequency tomography can be used for exploring
hyper-entanglement spatial qubit states of photon wave function and
for early disease detection in biophotonics.

\begin{acknowledgments}

\end{acknowledgments}


\newcommand{\noopsort}[1]{} \newcommand{\printfirst}[2]{#1}
  \newcommand{\singleletter}[1]{#1} \newcommand{\switchargs}[2]{#2#1}

\newpage

\pagebreak

\newpage




\end{document}